\documentclass[english,twocolumn,prl,amssymb,amsmath,superscriptaddress,showpacs]{revtex4}
\usepackage[latin9]{inputenc}
\usepackage{graphicx,color}
\usepackage[hidelinks]{hyperref}
\hypersetup{colorlinks=false}
\usepackage{babel}
\usepackage{braket}
\usepackage[super]{nth}

\usepackage{bbm}
\usepackage{amsfonts}
\usepackage{amssymb}
\usepackage{graphicx}
\usepackage{subfigure}
\usepackage{savesym}
\usepackage{amsmath}
\usepackage{txfonts}
\usepackage{multirow}
\usepackage{epstopdf}
\usepackage{lipsum}

\renewcommand{\Re}{\operatorname{Re}}

\newcommand{\be}{\begin{equation}}
\newcommand{\ee}{\end{equation}}
\newcommand{\bea}{\begin{eqnarray}}
\newcommand{\eea}{\end{eqnarray}}

\newcommand{\ue}{\ensuremath{{\rm{e}}}}
\newcommand{\Oz}{\ensuremath{{\mathcal{O}^z}}}

\newcommand{\spinup}{|\!\uparrow\rangle}
\newcommand{\spindown}{|\!\downarrow\rangle}

\begin{document}

\title{Spectroscopy of interacting quasiparticles in trapped ions}

\author{P. Jurcevic}
\affiliation{Institut f\"ur Quantenoptik und Quanteninformation,\"Osterreichische Akademie der Wissenschaften, Technikerstr. 21a, 6020 Innsbruck, Austria}
\affiliation{Institut f\"ur Experimentalphysik, Universit\"at Innsbruck, Technikerstr. 25, 6020 Innsbruck, Austria}

\author{P. Hauke}
\affiliation{Institut f\"ur Quantenoptik und Quanteninformation,\"Osterreichische Akademie der Wissenschaften, Technikerstr. 21a, 6020 Innsbruck, Austria}
\affiliation{Institut f\"ur Theoretische Physik, Universit\"at Innsbruck, Technikerstr. 25, 6020 Innsbruck, Austria}

\author{C. Maier}
\affiliation{Institut f\"ur Quantenoptik und Quanteninformation,\"Osterreichische Akademie der Wissenschaften, Technikerstr. 21a, 6020 Innsbruck, Austria}
\affiliation{Institut f\"ur Experimentalphysik, Universit\"at Innsbruck, Technikerstr. 25, 6020 Innsbruck, Austria}

\author{C. Hempel}
\affiliation{Institut f\"ur Quantenoptik und Quanteninformation,\"Osterreichische Akademie der Wissenschaften, Technikerstr. 21a, 6020 Innsbruck, Austria}
\affiliation{Institut f\"ur Experimentalphysik, Universit\"at Innsbruck, Technikerstr. 25, 6020 Innsbruck, Austria}

\author{B. P. Lanyon}
\affiliation{Institut f\"ur Quantenoptik und Quanteninformation,\"Osterreichische Akademie der Wissenschaften, Technikerstr. 21a, 6020 Innsbruck, Austria}
\affiliation{Institut f\"ur Experimentalphysik, Universit\"at Innsbruck, Technikerstr. 25, 6020 Innsbruck, Austria}

\author{R. Blatt}
\affiliation{Institut f\"ur Quantenoptik und Quanteninformation,\"Osterreichische Akademie der Wissenschaften, Technikerstr. 21a, 6020 Innsbruck, Austria}
\affiliation{Institut f\"ur Experimentalphysik, Universit\"at Innsbruck, Technikerstr. 25, 6020 Innsbruck, Austria}

\author{C. F. Roos}
\affiliation{Institut f\"ur Quantenoptik und Quanteninformation,\"Osterreichische Akademie der Wissenschaften, Technikerstr. 21a, 6020 Innsbruck, Austria}
\affiliation{Institut f\"ur Experimentalphysik, Universit\"at Innsbruck, Technikerstr. 25, 6020 Innsbruck, Austria}

\begin{abstract}

The static and dynamic properties of many-body quantum systems are often well described by collective excitations, known as quasiparticles. Engineered quantum systems offer the opportunity to study such emergent phenomena in a precisely controlled and otherwise inaccessible way. We present a spectroscopic technique to study artificial quantum matter and use it for characterizing quasiparticles in a many-body system of trapped atomic ions. Our approach is to excite combinations of the system's fundamental quasiparticle eigenmodes, given by delocalised spin waves. By observing the dynamical response to superpositions of such eigenmodes, we extract the system dispersion relation, magnetic order, and even detect signatures of quasiparticle interactions. Our technique is not limited to trapped ions, and it is suitable for verifying quantum simulators by tuning them into regimes where the collective excitations have a simple form. 

\end{abstract}
\date{\today}
\maketitle

The level of experimental control over engineered many-body quantum systems has been rapidly improving in recent years. Atoms in optical lattices \cite{Bloch2012} and ions in electrodynamic traps \cite{Blatt2012}, for example, now offer unprecedented access to the states, dynamics, and observables of interacting quantum systems. As these engineered systems become larger and more complex, conventional tomographic methods \cite{Haffner2005} for characterising their states and processes soon become either inefficient or of limited applicability \cite{Cramer:2010fk}. New methods are required to verify what has been built in the laboratory and to study the properties of these fascinating systems.

In their low-energy sector, interacting many-body systems are often well described in terms of collective excitations, with effective mass, dispersion relation, and scattering properties. Such emergent excitations can be understood as weakly interacting quasiparticles, which are responsible for a range of dynamical properties of the underlying system \cite{Sachdev2000}. Recently, experiments in systems of atoms and ions have demonstrated the central role that quasiparticles play in the transport of information and entanglement \cite{Cheneau2012,Fukuhara2013,Jurcevic2014}. The ability to precisely measure the properties of a system's quasiparticles, therefore, becomes a valuable tool for studying many-body quantum systems in the laboratory. 

\begin{figure}[t]
\begin{center}
\includegraphics[width=0.48\textwidth]{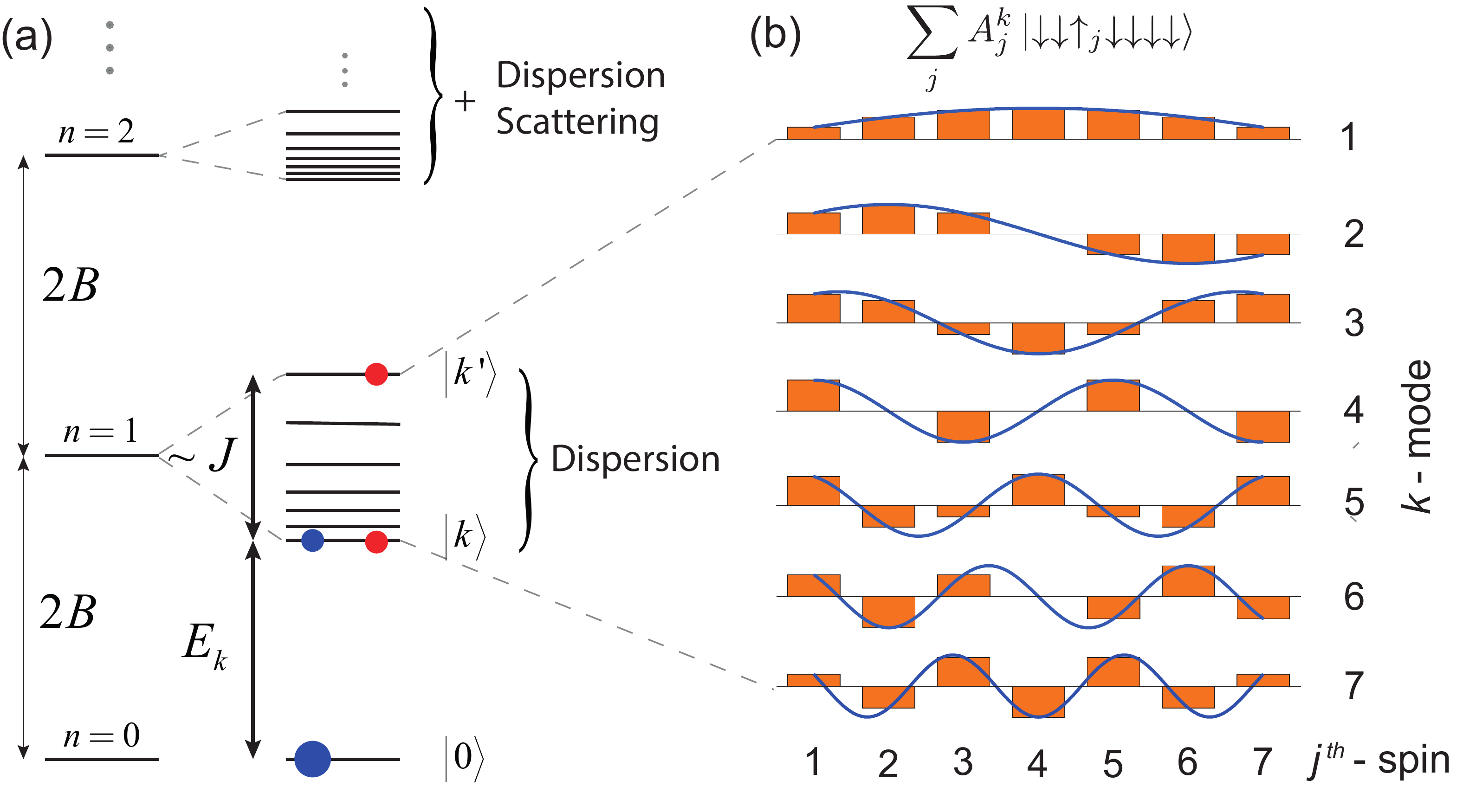}
\caption{\label{fig1} 
(a) Energy spectrum of model Hamiltonian Eq.~\eqref{eq:HXY}, for $B\gg J$ and 7 spins.
The degeneracy of subspaces with different excitation number $n$ is lifted by the spin--spin interactions $J$. 
The system response to superpositions $\ket{0}+\ket{k}$ (blue spheres) provides information about the absolute energy $E_k$ of spin-wave state $\ket{k}$. The response to superpositions $\ket{k}+\ket{k'}$ (red spheres) enables high-precision measurements of the system dispersion relation. Quantum dynamics in higher excitation manifolds ($n\geq 2$) involve both spin-wave dispersion and scattering. 
(b) Spin-wave amplitudes $A_j^k$. 
}
\end{center}
\end{figure}

Spectroscopy is an established approach to studying natural quantum systems and their emergent phenomena, well-established techniques including, e.g.,  photoemission spectroscopy in electronic systems \cite{Damascelli2004} and neutron scattering in magnetic materials \cite{Parker2013}. Currently, there is important theoretical \cite{Knap2013, Kitagawa2010,Dorner2013,Gessner2014} and experimental \cite{Stenger1999,Veeravalli2008,Ernst2010,Heinze2011,Britton2012,Yan2013,Langen2013, Senko2014,Landig2015} progress on developing spectroscopic techniques for engineered quantum systems. In this Letter, we present a spectroscopic technique for characterizing the low-lying energy spectrum of an engineered quantum many-body  system, and apply it to study emergent quasiparticles in a system of trapped atomic ions. Our approach exploits single-particle (-ion) control to excite individual quasiparticle modes. Measuring the system's dynamical response to superpositions of such single excitations, allows the absolute quasiparticle energies and system dispersion relation to be determined. Furthermore, by probing the response to superpositions of multi-quasiparticle states we are able to resolve shifts in the energy spectrum due to quasiparticle interactions. 

Our system is a 1D chain of $^{40}\mathrm{Ca}^+$ ions in a linear ion trap. In each ion $j=1\dots N$, two long-lived electronic states represent a quantum spin-1/2 particle with associated Pauli-matrices $\sigma^\beta_{j}$ ($\beta=x, y, z$). Under the influence of laser-induced forces, the system is ideally described by a one-dimensional transverse Ising model \cite{Porras2004, Supplementary2014}, with Hamiltonian
\begin{equation}
H=\hbar\sum_{i<j}J_{ij} \sigma_i^x\sigma_j^x +\hbar B \sum_j \sigma_j^z\,,
\label{eq:HXY}
\end{equation}
where $\sigma_j^\pm = (\sigma_j^x \pm i \sigma_j^y)$ are spin raising and lowering operators. The spin--spin coupling matrix $J_{ij}\sim J/|i-j|^\alpha$ approximately depends as a power-law on spin separation $|i-j|$, 
where $0<\alpha<3$ \cite{Jurcevic2014, Richerme2014}. All experiments presented here are carried out in a system of $N=7$ spins where $\alpha \approx 1.1$, corresponding to interactions of intermediate range \cite{Hauke2013}. 

We work with a strong transverse field $B\gg J$, in which case the energy spectrum of $H$ splits into $N+1$ subspaces, each containing eigenstates with the same integer `excitation number' $n\equiv \sum_j (\braket{\sigma_j^z}+1)/2$, i.e., total number of up-pointing spins, $\uparrow$ (see Fig.~\ref{fig1}(a) and \cite{Supplementary2014}). In this limit, $H$ conserves $n$. Therefore the subspaces are uncoupled and there is no population transfer between them during time dynamics. The eigenstates in the single-excitation subspace ($n=1$) are delocalised standing waves of spin excitation $|k\rangle=\sum_{j=1}^N A_j^k \ket{\uparrow_j} \otimes \ket{\downarrow}^{\otimes (N-1)}$, illustrated in Fig.~\ref{fig1}(b). These states $\ket{k}$ are the fundamental quasiparticle modes of the system, with eigenenergies $E_k$ that define the dispersion relationship.

At the basis of our spectroscopic method is the preparation of superpositions of eigenstates. 
Specifically, beginning with the ground state $\ket{\downarrow}^{\otimes N}{=}\ket{0}$, we use a laser, focused to a single ion and rapidly switchable between different ions, to rotate each spin $j$ into the state $\ket{\theta_j}\equiv\cos\left(\theta_j\right)\ket{\downarrow_j}+\sin\left(\theta_j\right)\ket{\uparrow_j}$. As we will show, preparing different superpositions allows us to spectroscopically measure energy gaps from the beat notes 
observed between eigenstates. 

\emph{Spectroscopy of ground-state gaps.---}
First, we measure the energies $E_k$ of the single-excitation eigenstates $\ket{k}$ relative to the absolute ground state. 
Although the $\ket{k}$ modes are delocalised, entangled states, we can prepare product states that are good approximations by setting  $\theta_j=\tan^{-1}\left(\gamma A_j^k\right)$, yielding the $N$-spin input state
\begin{equation}
\label{eq:initialState}
\ket{\psi_k} \propto\bigotimes_{j=1}^N \left[\ket{\downarrow_j}+\gamma A_j^k |\uparrow_j\rangle\right] = |0\rangle + \gamma \ket{k} + \mathcal{O}\left(\gamma^2\right)\,.
\end{equation}
Preparing this state requires no detailed \emph{a priori} knowledge of the Hamiltonian. As we will see, a strong signal is obtained when modelling the exact eigenmodes by generic standing waves with amplitudes \mbox{$A_j^k=\sqrt{\frac{2}{N+1}}\sin\left(\frac{kj\pi}{N+1}\right)$}. For open boundary conditions and nearest-neighbour interactions ($\alpha=\infty$), these form the exact eigenstates. In case of finite-range interactions ($\alpha<\infty$), the amplitudes $A_j^k$ are modified, but the number of wave-function nodes remains an accurate way to characterize the eigenstates. 

After preparing the system in $\ket{\psi_k}$, we let it evolve to \mbox{$\ket{\psi(t)}=\exp(-i Ht/\hbar)\ket{\psi_k}$}, and measure its single-spin magnetizations in the $x$-$y$ plane of the Bloch sphere. Specifically, we measure in a rotating frame where \mbox{$\tilde{x}=\cos(2Bt)x-\sin(2Bt)y$} and \mbox{$\tilde{y}=\sin(2Bt)x+\cos(2Bt)y$} \cite{Supplementary2014}. The corresponding expectation values are $\langle\sigma_j^{\tilde{x}(\tilde{y})}(t)\rangle=\bra{\psi(t)} \sigma_j^{\tilde{x}(\tilde{y})} \ket{\psi(t)} \approx 2 \gamma \,\Re[\exp(-i \varepsilon_k t/\hbar) \bra{0} \sigma_j^{\tilde{x}(\tilde{y})} \ket{k}] + \mathcal{O}\left(\gamma^3\right)$, where $\hbar=h/2\pi$ is the reduced Planck constant. These expectation values oscillate at a frequency determined by the energy shift $\varepsilon_k=E_k-2 B$ that result from the spin--spin couplings $J_{ij}$ [Fig.~\ref{fig1}(a)]. The {shift} $\varepsilon_k$ can then be extracted via a Fourier transform. To maximize the signal, we transform $M_{\tilde{x}}(t)+iM_{\tilde{y}}(t)$, where $M_{\tilde{x}(\tilde{y})}(t)=\sum_j A_j^k\langle\sigma_j^{\tilde{x}(\tilde{y})}(t)\rangle$, and mirror the result to negative times to minimize transform artifacts. 
For these measurements, we choose $\gamma$ such that the probability of creating more than one quasi-particle is only $p_2=0.02$, thus avoiding spurious frequency components in the Fourier transform due to energy gaps between states with one and two quasiparticles. 

\begin{figure}[tb]
\begin{center}
\includegraphics[width=0.48\textwidth]{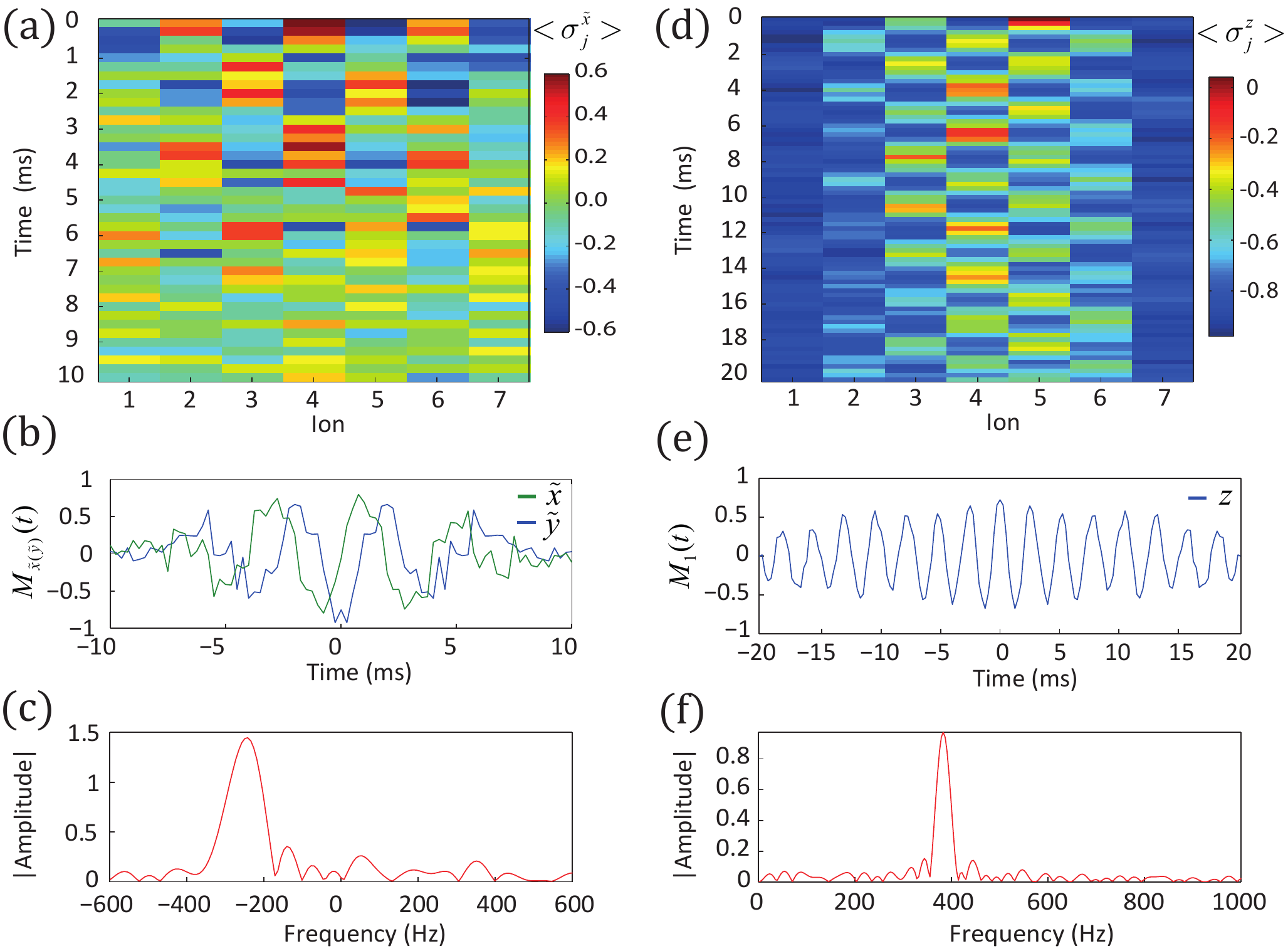}
\caption{\label{fig2}
Energy measurements of quasiparticle states. (a-c) An absolute energy probed with initial state $\ket{0}+\ket{k=7}$ and $\gamma=0.4$. (a) Spin-resolved $\tilde{x}$-magnetization dynamics. (b) Summed signals $M_{\tilde{x}(\tilde{y})}(t)$ along $\tilde{x}$ and $\tilde{y}$. (c) The Fourier transform of $M_{\tilde{x}}(t)+iM_{\tilde{y}}(t)$ shows a peak at $-249 \pm 42 ~\mathrm{Hz}$, revealing the downward energy shift of the $\ket{k=7}$ state caused by the spin--spin interactions.
(d-f) A relative energy probed with initial state $\ket{k=1}+\ket{k=7}$ and $\gamma=0.7$. (d) $z$-magnetization dynamics, obtained by post-selecting outcomes with exactly one spin-excitation. (e) Summed signal $M_1(t)$, see text. (f) The Fourier transform of $M_1(t)$ shows a peak at $383 \pm 13~\mathrm{Hz}$, revealing the energy splitting of the two quasiparticle states. In (a-c) $\gamma=0.4$, corresponding to a probability of having zero (one) excitation of $p_0=0.74$ ($p_1=0.24$). For (d-f) $\gamma=0.7$, yielding $p_0=0.43$ and $p_1=0.42$.
}
\end{center}
\vspace{-7mm}
\end{figure}

Results for $\ket{\psi_{k=7}}$ are presented in Fig.~\ref{fig2}(a-c). 
The Fourier transform reveals a distinct peak at $\varepsilon_{k=7}/h= -249 \pm 42 ~\mathrm{Hz}$, which reveals the interaction-induced dispersion $\varepsilon_k$ of mode $\ket{k}$ around the band center $2 B$. From the observation that $\varepsilon_{k=7}<0$ and the anti-aligned nearest-neighbour amplitudes of $\ket{k=7}$, we can infer that \mbox{$J_{ij}>0$}, i.e., the spin--spin couplings are antiferromagnetic. Analogous results are found for other modes. We observe, e.g., a positive energy shift for $\varepsilon_{k=1}$, in agreement with $J_{ij}>0$ and the requirement $\sum_{k=1}^{7} \varepsilon_{k}=0$. 
It is worth noting that the measured quantity $M_{\tilde{x}(\tilde{y})}(t)$ is equal to the quasiparticle Green's function $G_k(t)$, which contains all essential information about the quasiparticles, such as energies, scattering rates, and life times \cite{fetter, Knap2013}. 

\emph{Spectroscopy of the dispersion relation.---}
The measurement precision of the absolute energies $E_k$ is limited by dephasing due to laser-frequency and ambient magnetic-field noise. As subspaces with fixed excitation number $n$ are decoherence-free with respect to these noise sources \cite{Jurcevic2014}, the relative energies $\Delta E_{k,k'}=E_{k}-E_{k^\prime}$ 
can be measured with much higher resolution by preparing superpositions of the type $|k\rangle+|k^\prime\rangle$.

Here, we initialize the system in a product state similar to Eq.~\eqref{eq:initialState}, and choose the rotation angle  \mbox{$\theta_j = \tan^{-1}\left[\gamma \left( A_j^k + A_j^{k'} \right)\right]$} to create the state
\begin{equation}
\label{eq:psikkprime}
\ket{\psi_{kk^\prime}}= \ket{0} + \gamma \left(\ket{k} +\ket{k'}\right) + \mathcal{O}(\gamma^2).
\end{equation}
We then measure the time-evolved state in the $z$-basis. Since $H$ commutes with such measurements, this allows post-selection of the $\ket{k} +\ket{k'}$ contribution to the initial state by retaining only measurements projected on the single-excitation subspace. The individual $z$-magnetizations are added up to obtain the signal $M_1(t)=\sum_j c_j\langle\sigma_j^z(t)\rangle=\cos(\Delta E_{k,k'}t/\hbar)\sum_jc_j\langle k|\sigma_j^z|{k'}\rangle$ with \mbox{$c_j=\mathrm{sign}\left(A_j^k A_j^{k'}\right)$}. A Fourier transform of $M_1(t)$ yields the energy difference $\Delta E_{k,k'}$  between quasiparticle states $\ket{k}$ and $\ket{k^\prime}$. 

\begin{figure}[t]
\begin{center}
\includegraphics[width=0.48\textwidth]{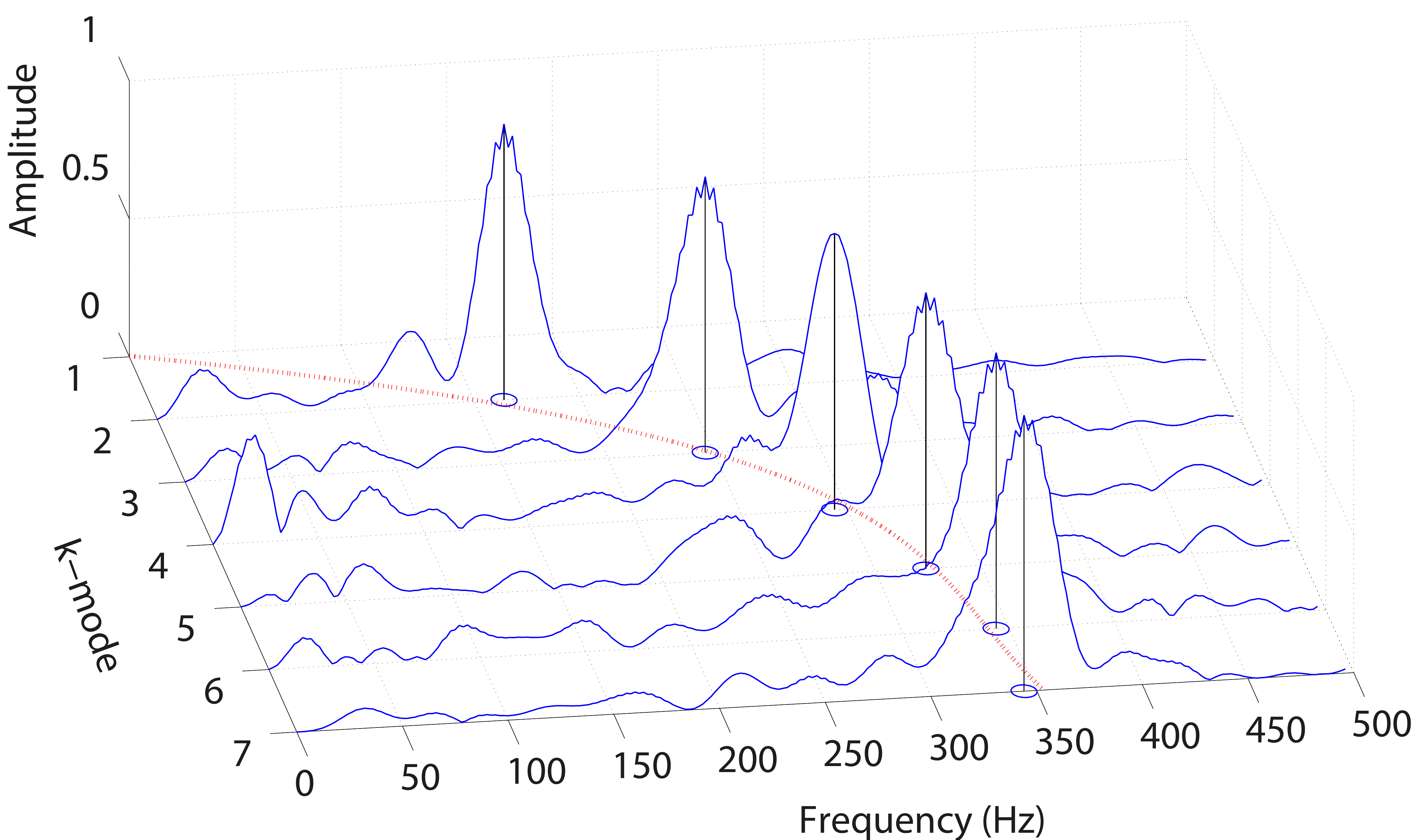}
\caption{\label{fig3} Quasiparticle dispersion relation, obtained from Fourier transforming the measured dynamics of the initial states $\ket{k=1}+\ket{k'}$ with $k^\prime=2\ldots 7$ (blue traces).
The blue circles mark the positions of peak maxima which closely match the theoretical dispersion relation (red line).
}
\end{center}
\end{figure}

Figure~\ref{fig2}(d-f) shows experimental data for a superposition of the quasiparticle states with lowest ($k'=7$) and highest ($k=1$) energies. Compared to \mbox{Figure~\ref{fig2}(a-c)}, the much longer coherence times within the $n=1$ subspace result in a significantly higher spectral resolution. By repeating this experiment for superpositions of different spin waves, we reconstruct the complete single-quasiparticle dispersion relation, measured relative to $k=1$. In Fig.\ \ref{fig3}, the results are compared with theoretical predictions, obtained by diagonalization of the model's coupling matrix $J_{ij}$, with $J$ as only free parameter. Instead of a cosine dispersion typical for short-range physics, the data reproduces very well the large group velocity ($\partial\omega/\partial k$) at $k=1$, as expected for long-range interactions with $\alpha<2$ \cite{Hauke2013,Cevolani2015}. The fact that each of the $N$ frequency spectra in Fig.~\ref{fig3} is dominated by a single peak demonstrates that the heuristically chosen sine-wave coefficients $A_j^k$ are, indeed, close to the ones of the exact eigenstates.

\emph{Spectroscopy of interacting quasiparticles.---}

Until now, we probed single quasiparticle dynamics.  
In the spectroscopy scheme based on initial states like Eq.~\eqref{eq:psikkprime}, quasiparticle excitation numbers above $n=1$ are naturally realized in higher orders of $\gamma$. We can therefore go further and perform spectroscopy in higher-excitation sectors by post-selecting time-evolved states with $n$ spin excitations. This provides an opportunity to characterize interactions between pairs of quasiparticle excitations.

\begin{figure}
\begin{center}
\includegraphics[width=0.48 \textwidth]{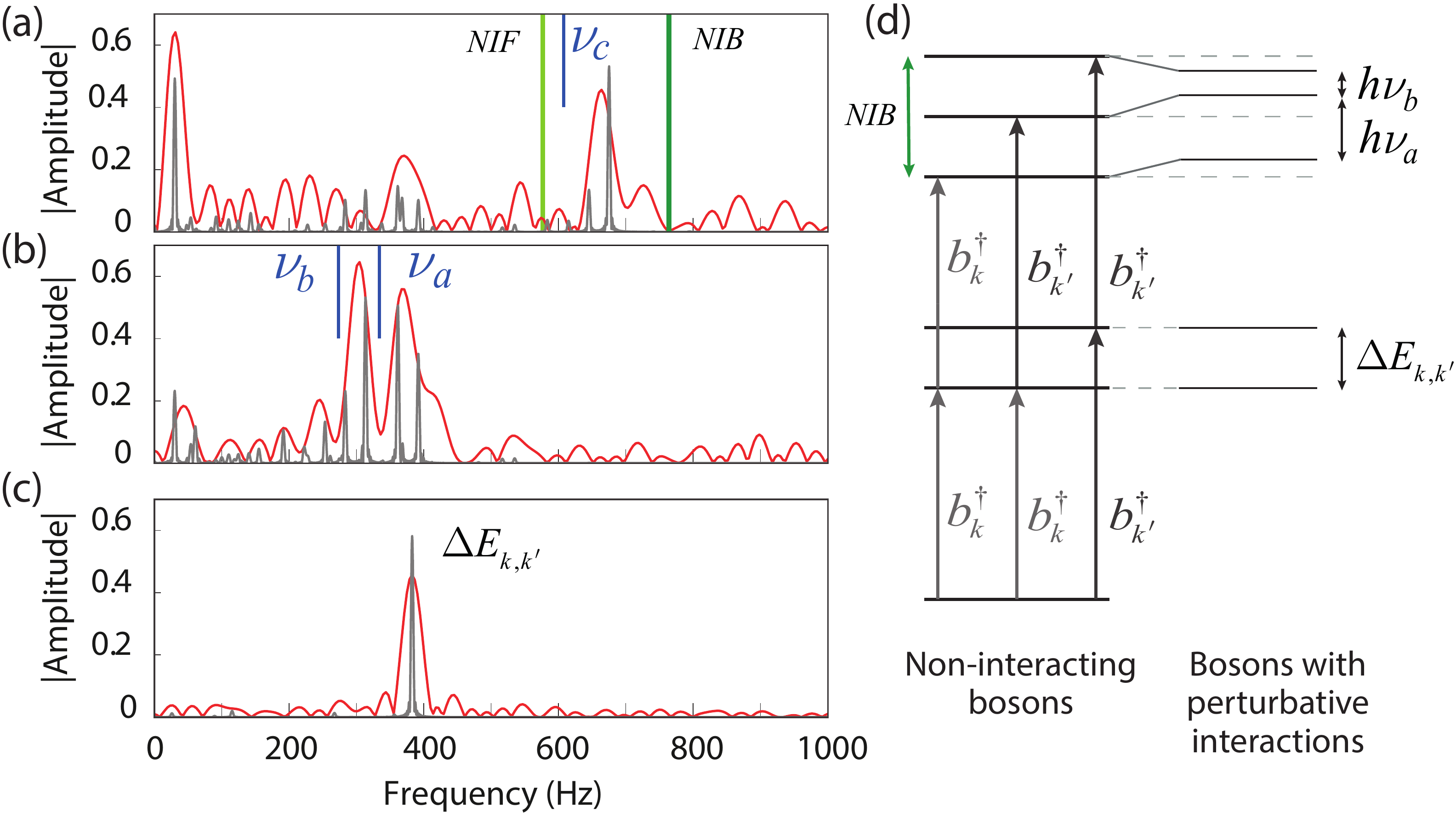}
\caption{\label{fig4} 
Spectroscopy of quasiparticle interactions. 
(a,b) Fourier spectra of response to the two-quasiparticle state $\ket{\psi^{(2)}_{kk^\prime=1,7}}$ with $\gamma=1.4$ (red curves), obtained from summed signals $M_{2a}(t)$ and $M_{2b}(t)$, respectively. \cite{Supplementary2014}. The highest frequency in (a) is inconsistent with the predictions for both non-interacting fermions (NIF) and non-interacting bosons (NIB), indicated by light, respectively dark, green lines. 
For NIB, e.g., one would expect a signal at twice the single-particle energy splitting $\Delta E_{k,k'}$, plotted in panel (c). 
The measured peak positions in (a,b) compare well with those of a longer simulated evolution, indicated by the  narrower gray curves. A perturbative treatment of quasiparticle interactions (blue solid half-lines) agrees qualitatively with the observed interaction shifts, but overestimates the effects. Here, $\gamma=1.4$ yielding $p_0=0.08, p_1=0.33$, and $p_2=0.41$ in order to maximize the two-quasiparticle population.
(d) Schematic energy diagram. Left: In a non-interacting system of free bosons, all higher excitation eigenstates can be inferred from the single-particle subspace by successive application of creation operators. Right: For interacting hard-core bosons, the energy levels acquire a density- and mode-dependent interaction shift. 
}
\end{center}
\vspace{-5mm}
\end{figure}

We now prepare product states $\ket{\psi_{k,k^\prime}}$, Eq.~\eqref{eq:psikkprime}, with larger $\gamma$, so that the next-order contribution $\ket{\psi^{(2)}_{kk^\prime}}=\gamma^2 \left[\sum_j\left( A_j^k + A_j^{k'} \right)\sigma_j^+\right]^2\equiv \gamma^2\left(\ket{kk}+2\ket{kk'}+\ket{k'k'}\right)$ becomes important. After evolving $\ket{\psi_{kk^\prime}}$ under $H$ as before,  we perform a projective measurement of  $\sigma_j^z(t)$. Afterwards, we post-select measurement outcomes with two excitations, compute the observables $\mathcal{P}_{ij}^z=\frac{1}{4} \langle(1+\sigma_i^z)(1+\sigma_j^z)\rangle$, and Fourier-transform the sum $M_{2a(b)}(t)=\sum_{i\neq j} c_{ij}^{a(b)}\mathcal{P}_{ij}^z(t)$ with suitable weights $c_{ij}^{a}$ or $c_{ij}^{b}$ \cite{Supplementary2014}. In parallel, we post-select outcomes with a single excitation and thereby measure $M_1(t)$ to obtain the single-particle energy splitting $|\Delta E_{k,k'}|/h$ as a reference [Fig.\ \ref{fig4}(c)]. 

Without interactions between quasiparticles, higher excited states would be simple combinations, determined by the quasiparticle statistics, of the single-excitation states. The measured frequency spectrum for $\ket{\psi^{(2)}_{k=1,k^{\prime}=7}}$, however, reveals that the quasiparticles do interact: the peak in Fig.\ \ref{fig4}(a) at $665 \pm 14$~Hz is significantly lower than $2|\Delta E_{k,k'}|/h = 765 \pm 32$~Hz, which would be expected for non-interacting bosons (NIB) from double population of the states $k$ and $k'$. Furthermore, the aforementioned peak is higher than the maximum allowed for non-interacting fermions (NIF), which is, due to Pauli exclusion, $|E_{k=7}+E_{k=6}-E_{k=1}-E_{k=2}|/h=585 \pm 21$~Hz. These measurements show that the quasiparticles in our system cannot be described by a simple free particle model, which is a consequence of the long-range nature of our interactions. A NIF model would be accurate only for strictly nearest-neighbour interactions \cite{Pfeuty1970}. 

For a qualitative understanding of the two-particle spectra, it is illustrative to map the spin operators to hard-core bosonic creation and annihilation operators~\cite{Holstein1940}, $\sigma_j^+ \rightarrow \tilde{b}_j^\dagger$ ($\sigma_j^- \rightarrow \tilde{b}_j$), with quasiparticle modes $|k\rangle=b_k^\dagger|0\rangle = \sum_j A_j^k \tilde{b}_j^\dagger|0\rangle$. Note that the two-particle states $\ket{kk'}=b_k^\dagger b_{k'}^\dagger \ket{0}$, $k^{(\prime)}=1\dots N$, are not eigenstates of the $n=2$ subspace; there would be $N(N+1)/2$ different states $\ket{kk'}$, whereas the $n=2$ subspace has only $N(N-1)/2$ dimensions because of the hard-core constraint imposed by having a spin-1/2 system. Nevertheless, for $N\gg 2$, we expect $\ket{kk'}$ to have a significant  overlap with a two-quasiparticle eigenstate.  

Without the hard-core constraint, $\ket{\psi^{(2)}_{kk^\prime}}$ would populate three exact eigenstates, whose energies would be the sums of the individual quasiparticle energies $E_k$ and $E_{k'}$ [see Fig.~\ref{fig4}(d)]. In that case, we would measure two distinct spectral peaks at frequencies $|\Delta E_{k,k'}|/h$ and $2|\Delta E_{k,k'}|/h$.  However, the hard-core interactions induce an interaction shift away from this value. For low quasiparticle densities, $n \ll N$, qualitative insights into interaction effects can be obtained by treating the hard-core interaction as a small perturbation. 
First-order perturbation theory predicts energy shifts $V_{k,k'}$ of the states $\ket{kk'}$ \cite{Supplementary2014}, and, consequently, beatnotes at $\nu_a=| \Delta E_{k,k'} + V_{k,k} - V_{k,k'}|/h$, 
$\nu_b=| \Delta E_{k,k'} - V_{k',k'} + V_{k,k'}|/h$, and $\nu_c=\nu_a+\nu_b$. In Fig.\ \ref{fig4}(a-b), the frequency spectrum for different summations of the data is shown for $\ket{\psi^{(2)}_{k,k^\prime=1,7}}$, where, indeed, we observe three distinct peaks near $\nu_a$, $\nu_b$ and $\nu_c$.

While the predictions of perturbation theory prove qualitatively correct, in terms of the existence and direction of interaction shifts, the effects are overestimated. In addition to the predicted three-peak structure, and their sum and difference frequencies, another large peak is observed at a very low frequency [Fig.\ \ref{fig4}(a)]. This peak can be explained by studying the exact eigenstates, where we find that the initial state has substantial overlap with a fourth eigenstate not expected from first-order perturbation theory. The reason for these discrepancies is that for the small chain considered the quasiparticle density, $n/N=2/7$, is not sufficiently low to quantitatively treat the hard-core interaction as a small perturbation. With increasing system size $n/N\rightarrow 0$, the quality of perturbation theory will improve and the states $\ket{kk'}$ should provide closer approximations to the true eigenstates of the two-excitation subspace of $H$. 

One can use this scheme to study interaction effects between different combinations of quasiparticle states. The resulting spectra can be much more complex than the one presented in Fig.~\ref{fig4}. For example, $\ket{\psi^{(2)}_{k,k^\prime=1,4}}$ has considerable overlap with many of the exact two-excitation eigenstates, which can be interpreted as a strong scattering of quasiparticles into other modes \cite{Supplementary2014}. 
This scattering reduces the quasiparticle lifetime, and induces broader spectral features. 

\emph{Conclusion.---} We realized a Ramsey-like technique for measuring the low-lying energy spectrum of an artificial quantum many-body system of trapped ions.  Our technique enabled a complete characterization of the system's quasiparticles, including the dispersion relationship and scattering properties. While the former determines how information is transported in our system, the latter shows that our system cannot be described by a simple model of non-interacting particles. The technique works by tuning the system into regimes where collective excitations have a simple form, allowing for many-body couplings to be precisely studied. Future work could study the regimes where $B$ and $J$ are of similar size. Here, the quasiparticle peak in the spin-wave Green's function would broaden and eventually disappear at the quantum critical point. Furthermore, it would be interesting to investigate whether multi-dimensional spectroscopy techniques developed for NMR or femtosecond spectroscopy \cite{Ernst1987, Mukamel2000} could be combined \cite{Schlawin2014} with the single-site addressability available in many artificial systems to measure finite quasiparticle lifetimes due to their  interactions or incoherent coupling to the environment. 

\begin{acknowledgments}
This work was supported by the Austrian Science Fund (FWF) under the grant number P25354-N20 and the SFB FoQuS (FWF Project No.\ F4016-N23), by the European Commission via the integrated project SIQS, and by the Institut f\"ur Quanteninformation GmbH.
\end{acknowledgments}


\bibliographystyle{apsrev}

\newpage

\renewcommand{\theequation}{S\arabic{equation}}
\setcounter{equation}{0}
\renewcommand{\thefigure}{S\arabic{figure}}
\setcounter{figure}{0}

\section{Supplemental Material}

\subsection{Derivation of a transverse Ising model}
This section provides a short derivation of how to realize a transverse field Ising model in a linear string of ions. The mechanism is based on transversally illuminating the ion string by a bichromatic laser field that off-resonantly couples to the lower and upper vibrational sidebands of the transition between the two electronic states encoding the pseudo-spin. 

We start by considering a monochromatic running-wave laser field coupling the internal electronic states of an ion to the $2N$ transverse motional modes of the $N$ ion string. We work
in an interaction picture with respect to the Hamiltonian $H_0=\frac{\hbar\omega_0}{2} \sum_i \sigma_i^z + \sum_m\hbar\omega_m a_m^\dagger a_m$  \cite{Leibfried2003}, where $\hbar\omega_0$ is the energy splitting between the two pseudospin states and $\hbar\omega_m$ the energy of phonon mode $m$ with annihilation (creation) operator $a_m$ ($a_m^\dagger$). The interaction Hamiltonian $H_I$ for a monochromatic light field reads 

\begin{eqnarray}\label{H_int}
H_I&=& \frac{\hbar}{2} \sum_{j=1}^N \sum_{m=1}^{2N} \Omega_j \bigg[ \sigma_j^+ e^{-i(\epsilon t-\Phi_L)}\left(1+i \eta_{j,m}(a_m e^{-i\omega_m t}+a_m^\dagger e^{i\omega_m t})\right)\nonumber\\
&&  + h.c. \bigg]\,,
\end{eqnarray}

where $\Omega_j$ is the Rabi-frequency of the laser at the $i$-th ion, $\Phi_L$ the laser phase, $\epsilon=\omega_L-\omega_0$ the detuning of the laser frequency from the atomic transition, and $\eta_{j,m}$ is the Lamb--Dicke factor, which is proportional to the displacement of the $j^{th}$ ion in the $m^{th}$ collective mode. The summations run over all $N$ ions and over all $2N$ transverse modes. The spin-flip operators are given by $\sigma_j^\pm=\sigma_j^x \pm i \sigma_j^y$, where $\sigma_j^{x,y,z}$ are the usual Pauli matrices. Note that Eqn.~(S1) is valid under the usual Lamb--Dicke approximation ($\eta_{j,m} \ll 1$), which is appropriate for our experiments. 

In order to achieve an effective spin--spin interaction, the red and the blue sidebands of all $2N$ modes are off-resonantly driven with a bichromatic light-field with symmetric detunings $\epsilon_\pm=\pm(\omega_M + \Delta)$.  Here, $\Delta$ denotes the detuning of the laser with respect to the blue sideband of the motional mode with the highest frequency $\omega_M$. If we perform a rotating-wave approximation to retain only the coupling of the frequency component with $\epsilon_+$ ($\epsilon_-$) to the upper (lower) motional sidebands, the resulting Hamiltonian is given by
\begin{eqnarray}\label{H_bic}
H_{\rm{bic}}=\frac{\hbar}{2} \sum_{j=1}^N \sum_{m=1}^{2N} \eta _{j,m}\Omega_j \left(\sigma_j^+ + \sigma_j^-\right)\left( a_m e^{i\Delta_m t} + a_m^\dagger e^{-i\Delta_m t} \right)
\end{eqnarray}
where $\Delta_m=\Delta+(\omega_M-\omega_m)$. In the limit of $|\eta _{j,m}\Omega_j|\ll\Delta_m$, the motional states are only virtually excited and serve to mediate an Ising interaction between the spins. This effective spin--spin interaction can be calculated from second-order perturbation theory, resulting in the effective Hamiltonian 
\begin{eqnarray}\label{H_ClassIsing}
H_{\rm{eff}}=\hbar\sum_{i,j} J_{ij}\left(\sigma_i^+ + \sigma_i^-\right)\left(\sigma_j^+ + \sigma_j^- \right)
=\hbar\sum_{i,j} J_{ij}\,\sigma_i^x\sigma_j^x
\end{eqnarray}  
with spin--spin coupling matrix
\begin{eqnarray}
J_{ij}=\frac{\Omega_i \Omega_j}{4} \sum_{m=1}^{2N}\frac{\eta_{i,m} \eta_{j,m}}{\Delta + \omega_M-\omega_m}\,.
\end{eqnarray}
Further details of the derivation can be found in references such as \cite{Kim2011} or \cite{Schneider2012}.

If the detuning of the bichromatic laser beam contains a slight asymmetry, $\epsilon_\pm=\pm(\omega_M + \Delta) + \delta$, $\delta\ll\Delta$, the raising and lowering operators appearing in Eqs.~(\ref{H_bic}), (\ref{H_ClassIsing}) obtain a time-dependent phase factor, $\sigma^+\rightarrow\sigma^+e^{-i\delta t}$ and  $\sigma^-\rightarrow\sigma^-e^{i\delta t}$, and thus 
\begin{eqnarray}\label{H_timedependent}
H_{\rm{eff}} &=& \hbar \sum_{i,j}  J_{ij}\left(\sigma^+_i\sigma^+_je^{-2i\delta t} +\sigma^+_i\sigma^-_j +\sigma^-_i\sigma^+_j + \sigma^-_i\sigma^-_je^{2i\delta t}\right).\nonumber\\
\label{ss}
\end{eqnarray}

Additional terms $\propto\sigma_j^z a_m^\dagger a_m$ that arise from off-resonant excitation of the sideband transitions are smaller by a factor $\delta/\Delta$ and can be neglected in the limit $\delta\ll\Delta$ in which we work here. 
 
The time-dependent Hamiltonian in Eq.~(\ref{H_timedependent}) can be transformed into a time-independent Hamiltonian by writing $H_{\rm{eff}}=\left(H_{\rm{eff}}-H_\delta\right)+H_\delta$ and changing into an interaction picture with respect to $H_\delta=-\hbar\sum_j\frac{\delta}{2}\sigma_j^z$, generated by the unitary operator $U_{\delta}=e^{-iH_{\delta} t /\hbar}$,  
\begin{eqnarray}
H'_{\rm{eff}} &=& U_{\delta}^\dagger (H_{\rm{eff}}-H_\delta)U_{\delta}\nonumber\\
		&=&\hbar \sum_{i,j} \ J_{ij}~ \sigma^x_i\sigma^x_j +\hbar B \sum_j \sigma_j^z
		= H_{\rm{Ising}}\,.\label{H_Ising}
\end{eqnarray}
This Hamiltonian describes an Ising model in the effective transverse field $B=-\frac{\delta}{2}$. 

Our experiments are carried out in the regime $\delta \gg \max_{i,j} \{|J_{ij}|\}$, i.e., that of large transverse field. In that case, the phases of the double spin-flip terms $\sigma^+_i\sigma^+_j$ and $\sigma^-_i\sigma^-_j$ in Eqn. (\ref{ss}) are fast rotating and can be eliminated. Finally, moving into the same interaction picture as that used for Eq.~\eqref{H_Ising} one recovers the XY-Hamiltonian 

\begin{eqnarray}\label{H_XY}
H'_{\rm{eff}} &=& \hbar \sum_{i,j} \ J_{ij} \left( \sigma^+_i\sigma^-_j + h.c.\right) + \hbar B\sum_j \sigma_j^z = H_{XY}.
\end{eqnarray}

\subsection{Spectroscopic signals for single quasiparticle states}

We perform spectroscopy of a trapped-ion system modelled by the $XY$ Hamiltonian,  Eq.~(\ref{H_XY}), of $N$ coupled spins.  
As discussed in Fig.~1 in the main text, the spectrum of $H_{XY}$ splits into uncoupled subspaces with integer total excitation number (number of up-pointing spins), allowing us to study each subspace individually or to compare the energy of states within one subspace against the energy of a reference state in another subspace.
In the single-excitation subspace, one can calculate the eigenenergies ($E_k$) and eigenmodes ($\ket{k}={\sigma}_k^+\ket{0}$, with ${\sigma}_k^+=\sum_j A_{j}^k \sigma_j^z$) from the model Hamiltonian simply by diagonalizing the $N\times N$ matrix $J_{ij}$.

To spectroscopically measure the energies $E_k$, we use superpositions of eigenstates. These can be approximated by initial product states like 
\begin{subequations}
\bea
\ket{\psi_0}&=&\bigotimes_{j=1}^N \left[\cos(\theta_j) \ket{\downarrow}_j + \sin(\theta_j) \ket{\uparrow}_j\right] \\
		 &=&C \bigg[ 1 + \sum_j \tan(\theta_j) \sigma_j^+ + \sum_{i,j} \tan(\theta_i) \tan(\theta_j) \sigma_i^+ \sigma_j^+ \nonumber\\
		 & &+\mathcal{O}(\tan(\theta)^3) \bigg] \ket{0} \nonumber\,. 
\eea
\end{subequations}
Here, $C\equiv\prod_{j=1}^N \cos(\theta_j)$ provides the overall normalization of the wave function. \\

\subsection{Absolute quasiparticle energies}

For the spectroscopy of absolute energy values $E_k$, we choose $\tan(\theta_j) =\gamma A_{j}^{k}$ to generate the initial state 
\be
\ket{\psi_k} = C \left[ \ket{0} +\gamma \sum_j A_{j}^{k} \sigma_j^+ \ket{0}+ \mathcal{O}(\gamma^2) \right]  \,. 
\ee
The time-evolved state is then $|\psi(t)\rangle = \ue^{-i H_{XY} t}|\psi_k\rangle$. If this state is measured directly in the $x$-basis, with respect to the frame of Eqn. (\ref{H_XY}), the signal would read:
\begin{subequations}
\bea
\bra{\psi(t)}\sigma^x_i\ket{\psi(t)}&=&\bra{\psi_k}\ue^{i H_{XY} t} \,\sigma^x_i\, \ue^{- i H_{XY} t} \ket{\psi_k} \\
&=& C^2 \gamma \left(\ue^{- i E_k t/\hbar} \bra{0} \sigma^x_i \sigma_{k}^+ \ket{0} +\mathrm{c.c.} \right)+  \mathcal{O}(\gamma^3) \nonumber \\
&=& C^2 \gamma \left(\ue^{- i E_k t/\hbar} A_i^k +\mathrm{c.c.} \right)+  \mathcal{O}(\gamma^3) \nonumber
\eea
\label{ttt}
\end{subequations}
This observable contains a strong oscillation at the quasiparticle energy $E_k$. The contributions from individual spins (ions) can be constructively summed to $\sum_i A_i^k \bra{\psi(t)}\sigma^x_i\ket{\psi(t)} = \bra{0}\sigma^-_k(t) \sigma_k^+ \ket{0} +\mathrm{c.c.} +  \mathcal{O}(\gamma^3)$, with $\sigma^-_k(t)=\ue^{i H_{XY} t} \,\sigma^-_k\, \ue^{- i H_{XY} t} $. Note, to leading order this is equivalent to the single-particle Greens function \cite{Fetter1971}. 

However, we do not measure directly in the $x$-basis (or $y$-basis) of Eq.~(\ref{H_XY}), as assumed in Eq.~(\ref{ttt}). 
Measurements of the magnetization in the $x$-$y$ plane of the Bloch sphere are carried out by first performing $\pi/2$ pulses with a laser resonant with the pseudo-spin transition followed by a fluorescence measurement projecting the spin onto either $\spinup$ or $\spindown$. This corresponds to measurement of $\langle\sigma_x\rangle$ or $\langle\sigma_y\rangle$ in the reference frame of the Hamiltonian (\ref{H_bic}). When changing into the reference frame of the Hamiltonians (\ref{H_Ising}) or (\ref{H_XY}), these single-spin measurements correspond to measurements of 
\begin{eqnarray}
\sigma^{\tilde{x}} &=& e^{i\frac{\delta}{2}t\sigma^z} \sigma^x e^{-i\frac{\delta}{2}t\sigma^z}=\cos(\delta t)\sigma^x - \sin(\delta t)\sigma^y\\
\sigma^{\tilde{y}} &=& e^{i\frac{\delta}{2}t\sigma^z} \sigma^y e^{-i\frac{\delta}{2}t\sigma^z}=\sin(\delta t)\sigma^x + \cos(\delta t)\sigma^y\\
\sigma^{\tilde{z}} &=& e^{i\frac{\delta}{2}t\sigma^z} \sigma^z e^{-i\frac{\delta}{2}t\sigma^z}=\sigma^z \,.
\end{eqnarray}

Note that measurements in the logical ($z$-basis) are unaffected by this frame change. The consequence of measuring in a rotated frame in the $x$ or $y$ basis is simply that we observe oscillations due to energy gaps $\epsilon_k=E_k-2B$, as presented in Fig.2 (a-c) in the main text.

\subsubsection{Relative quasiparticle energies}

To observe beatnotes between different eigenenergies within the single-excitation subspace, we prepare approximate superpositions of two eigenfunctions, $\tan(\theta_j) =\gamma\left( A_{j}^{k_1}+ A_{j}^{k_2}\right)$, yielding the initial state 
\be
\ket{\psi_0}=C \left[ 1 +\gamma ({\sigma}_{k_1}^++{\sigma}_{k_2}^+) +\gamma^2 ({\sigma}_{k_1}^++{\sigma}_{k_2}^+)^2 +\mathcal{O}(\gamma^3) \right] \ket{0} \,.
\ee

When measuring the observable $\sigma_i^z$ for spin $i$ in the time evolved state $|\psi(t)\rangle = \ue^{-i H_{XY} t}|\psi_0\rangle$, the observed signal is 
\begin{widetext}
\begin{subequations}
\bea
\bra{\psi(t)}\sigma^z_i\ket{\psi(t)}=\bra{\psi_0}\ue^{i H_{XY} t} \,\sigma^z_i\, \ue^{- i H_{XY} t} \ket{\psi_0} =
& & C^2 \bigg[\bra{0} \sigma^z_i  \ket{0}\nonumber\\
& & +  \gamma^2\bra{0}\Big({\sigma}_{k_1}^-\ue^{iE_{k_1}t/\hbar}+{\sigma}_{k_2}^-\ue^{iE_{k_2}t/\hbar}\Big) \,\sigma^z_i  \Big({\sigma}_{k_1}^+\ue^{-iE_{k_1}t/\hbar}+{\sigma}_{k_2}^+\ue^{-iE_{k_2}t/\hbar}\Big) \ket{0}  \nonumber\\
& & +\,\gamma^4 \bra{0}\Big({\sigma}_{k_1}^-+{\sigma}_{k_2}^-\Big)^2 \ue^{i H_{XY} t} \,\sigma^z_i \ue^{-i H_{XY} t} \Big({\sigma}_{k_1}^++{\sigma}_{k_2}^+\Big)^2 \ket{0} \,\,\,\,+\mathcal{O}(\gamma^6)  \bigg]\,.
\eea
\end{subequations}
\end{widetext}

When measuring $\sigma^z_i$, post-selecting measurement outcomes with a fixed number of total spin-up excitations ($n$) allows the dynamics (and therefore energy gaps) in different subspaces to be studied independently. We define projection operators, $\Pi_n$, which denotes the projector onto the subspace with excitation number $n$. To study the single excitation subspace we extract the observable 
\bea
\bra{\psi(t)}\Pi_1\,\sigma^z_i\,\Pi_1\ket{\psi(t)} &=& C^2 \gamma^2\bigg[ \bra{k_1}\sigma^z_i\ket{k_1} \\ 
 &&+\bra{k_2}\sigma^z_i\ket{k_2}\nonumber \\
 &&+ 2\Re \Big( \ue^{i(E_{k_1}-E_{k_2})t/\hbar} \bra{k_1}\sigma^z_i\ket{k_2} \Big)\bigg]\nonumber\,,
\eea
Here, the signal oscillates with a frequency given by the energy difference of the eigenmodes. 
For the initial state where $\tan(\theta_j) =\gamma\left( A_{j}^{k_1}+ A_{j}^{k_2}\right)$, the oscillations have amplitude proportional to $2 A_j^{k_1} A_j^{k_2}$ (for $k_1\neq k_2$). 
The average signal from various spins is therefore maximized by Fourier-transforming the weighted sum  

\begin{subequations}
\bea
M_1(t)=\sum_j \mathrm{sign}\left(A_j^{k_1} A_j^{k_2}\right)\langle\Pi_1\sigma_j^z\Pi_1\rangle. 
\eea
\end{subequations}

\subsection{Spectroscopic signals for two quasiparticle states}

The dynamics in higher excitation subspaces becomes more complicated because the quasiparticle modes that diagonalize $H_{XY}$ in the single-excitation subspace are subject to the hard-core constraint $\sigma_j^+ \sigma_j^+ =0$. 
In a dilute system, i.e., when there are few excitations, $\sum_j \sigma_j^+\sigma_j^- /N \ll 1$, the excitations become approximately independent. We can then construct a perturbation theory around the non-interacting eigenmodes without any hard-core constraint. 

\subsubsection{Mapping spin operators to bosonic operators}
A formal way to construct a perturbation theory around the non-interacting eigenmodes is by applying the Holstein--Primakoff transformation \cite{Holstein1940} to obtain bosonic creation and annihilation operators $\tilde{b}_i^\dagger$ and $\tilde{b}_i$, 
\begin{subequations}
\begin{eqnarray}
	\sigma_j^-&\to& \sqrt{2S}\sqrt{1-\frac{\tilde{b}_j^\dagger \tilde{b}_j}{2S}}\tilde{b}_j \,,\\
	\sigma_j^+&\to& \sqrt{2S}\tilde{b}_j^\dagger \sqrt{1-\frac{\tilde{b}_j^\dagger \tilde{b}_j}{2S}}\,, \\
	\sigma_j^z&\to& -S+\tilde{b}_j^\dagger \tilde{b}_j \,, 	
\end{eqnarray}
\end{subequations}
where $S=\frac{1}{2}$ is the length of the spin (we keep the factor $2S$ as it provides a convenient way to distinguish different perturbative contributions). The Holstein--Primakoff transformation conserves the bosonic commutation relations between spins on different sites 
$[\sigma_i^-,\sigma_j^+]=0$, $i\neq j$, and the onsite hard-core constraint $\sigma_j^+\sigma_j^+=0$. In linear spin-wave theory (LSWT) \cite{Majlis}, one expands the spin Hamiltonian to leading order in the bosonic occupations, which amounts to neglecting the constraint by setting $\sqrt{1-\frac{\tilde{b}_j^\dagger \tilde{b}_j}{2S}}\to 1$. The result is the approximation
\be
\label{eq:HLSWT}
	H_{XY} \approx H_{\mathrm{LSWT}}= 2S \sum_{i<j}J_{ij}\left(\tilde{b}_i^\dagger \tilde{b}_j+h.c.\right) \,,
\ee
It is straightforward to diagonalize the linear spin-wave Hamiltonian $H_{\mathrm{LSWT}}$ to find its eigenenergies $E_k$ and eigenmodes $b_k^\dagger = \sum_j A_j^k \tilde{b}_j^\dagger$ (the spin waves), which reproduce the exact physics in the single-excitation subspace. If there was no hard-core constraint, combinations of spin waves would also define the eigenstates in higher occupation subspaces. For low spin-wave densities, one can expect this constraint to only play a perturbative role, motivating the use of Hamiltonian $H_{\mathrm{LSWT}}$ as starting point for a perturbative expansion in the spin-wave interactions. 

\subsubsection{Perturbative treatment of spin-wave interactions}
One can obtain leading corrections due to hard-core interactions by expanding the square roots to next order in the spin-wave density $\braket{{\tilde{b}_j^\dagger \tilde{b}_j}}/{2S}$, yielding 
\be
\label{eq:H2ndSWT}
	H_{XY} \approx H_{\mathrm{LSWT}} + \hat{V} \,,   
\ee
with $\hat{V}=-\frac 1 2 \sum_{i,j}J_{ij}\left(\tilde{b}_j^\dagger \tilde{b}_j^\dagger \tilde{b}_j \tilde{b}_i + \tilde{b}_j^\dagger \tilde{b}_i^\dagger \tilde{b}_i \tilde{b}_i\right)$. 
The eigenstates of the unperturbed system are given by $\ket{k_1,k_2}_{\mathrm{SW}}=b_{k_1}^\dagger b_{k_2}^\dagger\ket{0}/\sqrt{1+\delta_{k_1,k_2}}$, where $\ket{0}=\ket{\downarrow}^{\otimes N}$ is the spin-wave vacuum and the $\delta_{k_1,k_2}$ takes care of correct normalisation. 
The perturbative shifts of the corresponding energies are in first order 
\begin{equation}
V_{k_1,k_2} \equiv \bra{k_1,k_2} \hat{V} \ket{k_1,k_2}_{\mathrm{SW}} = -\frac{2 (E_{k_1}+E_{k_2}) {\cal M}_{k_1k_2}^{k_1k_2}}{1+\delta_{k_1k_2}}\,.
\end{equation}
The strength of the interactions is determined by the geometrical overlap between different spin-wave modes, 
\be
	{\cal M}_{k_1,k_2}^{k_3,k_4}= \sum_j A_j^{k_1} A_j^{k_2} A_j^{k_3} A_j^{k_4}\,, 
\ee
and can be visualised as a contact interaction vertex between two incoming and two outgoing waves. In this picture, the factor $1/2$ in the case $k_1=k_2$ can be understood from the fact that for indistinguishable waves there is only one outgoing channel, whereas there are two for distinguishable waves.  

\subsubsection{Expected signal}
We can use these approximate energies to evaluate the expected signal in the two-excitation manifold. To this end, we approximate the time evolution as 
\be
	\ue^{-i H_{XY} t/\hbar} \ket{k_1,k_2} \approx \ue^{-i (H_{\mathrm{LSWT}} +\hat{V}) t /\hbar} (\ket{k_1,k_2}_{\mathrm{SW}}+\ket{\delta\psi(k_1,k_2)})\,.
\ee
If we are interested only in the strongest frequencies, we can neglect corrections $\ket{\delta\psi(k_1,k_2)}$ to the states $\ket{k_1,k_2}=\sigma_{k_1}^+\sigma_{k_2}^+\ket{0}$, which correspond to scattering into other spin-wave modes, and approximate 
\be
	\ue^{-i H_{XY} t/\hbar} \ket{k_1,k_2} \approx \ue^{-i (E_{k_1}+E_{k_2} +V_{k_1,k_2}) t /\hbar} \ket{k_1,k_2}\,.
\ee
  
We now define $\Delta E_{k_1,k_2}=E_{k_1}- E_{k_2}$. For any observable $\Oz$ that is diagonal in the $\sigma^z$ Pauli matrices, the observed signal in the two-excitation subspace is 
\bea
& &\bra{\psi(t)}\Pi_2\,\Oz\,\Pi_2\ket{\psi(t)}\approx \nonumber \\
& & C^2\gamma^4 \left[ 
\bra{k_1k_1} \Oz\ket{k_1k_1} +\bra{k_2k_2} \Oz \ket{k_2k_2} + 4\bra{k_1k_2} \Oz \ket{k_1k_2}
\right. \nonumber\\
& & \left. \quad +2\cos\left(\frac{2 \Delta E_{k_1,k_2} + V_{k_1,k_1} - V_{k_2,k_2}}{\hbar} t\right)\bra{k_1k_1} \Oz \ket{k_2k_2} \right.\nonumber\\
& & \left. \quad +4\cos\left(\frac{\Delta E_{k_1,k_2} + V_{k_1,k_1} - V_{k_1,k_2} }{\hbar} t\right)\bra{k_1k_1} \Oz \ket{k_1k_2} \right.\nonumber \\
& & \left. \quad +4\cos\left(\frac{\Delta E_{k_1,k_2} + V_{k_1,k_2} - V_{k_2,k_2} }{\hbar} t\right)\bra{k_1k_2} \Oz \ket{k_2k_2} \right.\nonumber \\
& & \left. \quad + \mathrm{\quad corrections\quad from\quad other\quad states \quad}\right]\,.
\eea

This formula neglects all higher-order contributions in $\hat{V}$, including scatterings into other than the initially prepared modes $k_1,k_2$. 
As long as perturbation theory is a good description of the two-excitation subspace, we thus expect predominantly three frequencies in the time-evolution of the magnetization signal, which lie at 
$h \nu_a=| \Delta E_{k_1,k_2} + V_{k_1,k_1} - V_{k_1,k_2}|$, 
$h \nu_b=| \Delta E_{k_1,k_2} + V_{k_1,k_2} - V_{k_2,k_2}|$, and 
$h \nu_c=| 2\Delta E_{k_1,k_2} + V_{k_1,k_1} - V_{k_2,k_2}|$. 
Indeed, this three-peak structure with $\nu_a+\nu_b=\nu_c$ is what we observe in the data for $\left\{k_1=1,\, k_2=7\right\}$, as presented in Fig.~4 of the main text. 

To derive simple quantitative estimates, we use sine-wave amplitudes for the eigenfunctions, $A_{j}^k=\sqrt{\frac{2}{N+1}}\sin(k\cdot j \frac{\pi}{N+1})$, yielding the interaction-shifted frequencies   
$\nu_a = ( 1 - 0.113 ) \Delta E_{k_1,k_2}/h$, $\nu_b = ( 1 - 0.262 ) \Delta E_{k_1,k_2}/h$, and $\nu_c = ( 2 - 0.375 ) \Delta E_{k_1,k_2}/h$ for $\left\{k_1=1,\, k_2=7\right\}$. 

\subsubsection{Summation of spin-spin correlation measurements}
In the two-particle subspace, we study weighted sums of two-spin projectors $\mathcal{P}_{ij}^z=\frac 1 4 \left(\sigma_i^z +1 \right)\left(\sigma_j^z +1\right)$, i.e., the probability that both ions $i$ and $j$ are in the state $\ket{\uparrow}$. 
The corresponding expectation values are
\bea
& & \frac 1 4 \bra{k_1k_2} \left(\sigma_i^z \sigma_j^z + \sigma_i^z + \sigma_j^z +1\right) \ket{k_3k_4}=\nonumber \\
& &\left(A_i^{k_1}A_j^{k_2}+A_i^{k_2}A_j^{k_1}\right)\left(A_i^{k_3}A_j^{k_4}+A_i^{k_4}A_j^{k_3}\right)\,.
\eea

For the component with oscillation frequency $\nu_c$, we have $\frac 1 4 \bra{k_1k_1} \left(\sigma_i^z \sigma_j^z + \sigma_i^z + \sigma_j^z +1\right) \ket{k_2k_2}=4A_j^{k_1}A_i^{k_1} A_i^{k_2}A_j^{k_2}$, so that the signal is maximised by
\bea
M_{2a}(t)=\sum_{l<m}\mathrm{sign}\left(A_j^{k_1} A_i^{k_1} A_j^{k_2} A_i^{k_2} \right) \langle \mathcal{P}_{ij}^z \rangle\,.
\eea
For $\left\{k_1,k_2\right\}=\left\{1,7\right\}$, this choice simultaneously suppresses the expectation values associated to the other two oscillation frequencies, so that it allows one to cleanly extract a single one of the three expected Fourier components. 

Similarly, the component oscillating with $\nu_a$ has an amplitude $2|A_i^{k_1}|^2 A_j^{k_1} A_i^{k_2}+2|A_j^{k_1}|^2 A_i^{k_1} A_j^{k_2}$. If we use 
\bea
M_{2b}(t)=\sum_{i<j} \left(\mathrm{sign}\left(A_i^k A_j^{k'}\right) + \mathrm{sign}\left(A_i^{k} A_j^{k'} \right) \right)\langle \mathcal{P}_{ij}^z \rangle\,,
\eea
the signals due to this component and the one at $\nu_b$ are maximized while the one at $\nu_c$ gets suppressed. 

\subsubsection{Validity of first-order perturbation theory: additional experimental data}
\begin{figure}
\begin{center}
\includegraphics[width=0.48 \textwidth]{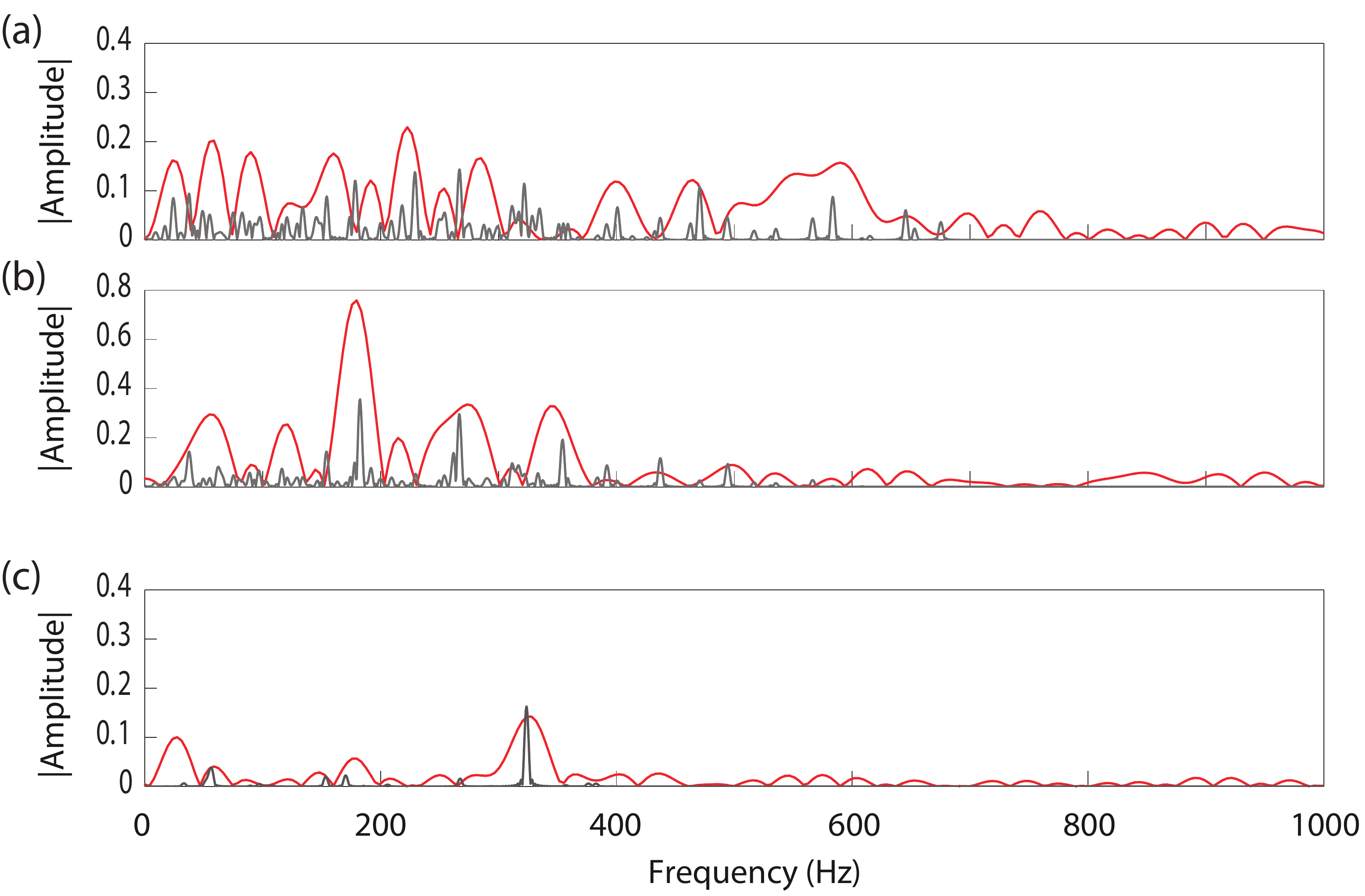}
\caption{\label{figsupp1} 
Spectroscopy of quasiparticle interactions.  (a,b) Fourier spectra of the two-quasiparticle state $\ket{\psi_{k,k'=1,4}^{(2)}}$, obtained from summed signals (red curves), with $M_{2a}(t)$ to enhance $\nu_c = \nu_a + \nu_b$ (panel a) and $M_{2a}(t)$ to increase signals at $\nu_{a,b}$ (panel b). 
The measured peak positions compare well with those of a simulated evolution with longer times, yielding a narrower bandwidth (grey curves). However due to strong quasiparticle scattering a large number of frequencies components becomes apparent. The weakly interacting single-particle eigenmodes no longer provide a good basis for describing the system. }
\end{center}
\end{figure}

The quality of the first-order perturbation theory for given eigenstates can be estimated by considering how well a state ${\ket{k_1k_2}}$ remains localized in the two-excitation state space when taking the hard-core constraint into account. 
This amounts to acting on the vacuum $\ket{0}$ with $b_{k_1}^\dagger b_{k_2}^\dagger$ and computing the overlap of the resulting state to the exact eigenstates. 
The state ${\ket{k_1=1,k_2=7}}$, constructed from single-particle states at the edge of the spectrum, for example, remains well localized in few exact eigenstates, and therefore qualitative predictions of simple first-order perturbation theory in the spin-wave interaction work quite well. Other combinations, on the other hand, such as ${\ket{k_1=1,k_2=4}}$ (Fig. \ref{figsupp1}) show many other spectral features deviating from the three-peak structure predicted by the perturbation theory. This comes from a considerable overlap with many of the true eigenstates in the two-excitation subspace, resulting in a diminished reliability of perturbation theory. 

Since the spin-wave scattering is proportional to the overlaps ${\cal M}_{k_1,k_2}^{k_3,k_4}$, which decrease with reduced excitation density, estimates from perturbation theory will improve with increasing system size (if the excitation number is kept fixed). For the small system considered, the overlaps are quite large, lying in the range ${\cal M}_{k_1,k_2}^{k_3,k_4}\approx 0.1-0.25$, which explains the lack of quantitative reliability of non-linear spin-wave theory. For $N\to \infty$, however, corrections such as ${\cal M}_{k_1,k_2}^{k_3,k_4}$ go to $0$ as $1/N$, and the states $\ket{k_1k_2}_{\mathrm{SW}}$ provide accurate approximations to the true eigenstates of the two-excitation subspace of $H_{XY}$.

\end{document}